\documentclass[10pt,letterpaper]{article}
\usepackage[top=0.85in,left=2.75in,footskip=0.75in]{geometry}

\usepackage{amsmath,amssymb}

\usepackage{changepage}

\usepackage[utf8x]{inputenc}

\usepackage{textcomp,marvosym}

\usepackage{cite}

\usepackage{nameref}

\usepackage{microtype}
\DisableLigatures[f]{encoding = *, family = * }
\usepackage[table]{xcolor}

\usepackage{array}

\usepackage{hyperref}

\newcolumntype{+}{!{\vrule width 2pt}}

\newlength\savedwidth

\raggedright
\usepackage[aboveskip=1pt,labelfont=bf,labelsep=period,justification=raggedright,singlelinecheck=off]{caption}

\makeatletter
\renewcommand{\@biblabel}[1]{\quad#1.}
\makeatother

\usepackage{lastpage,fancyhdr,graphicx}
\usepackage{epstopdf}
\fancyheadoffset[L]{2.25in}
\fancyfootoffset[L]{2.25in}
\IfFileExists{url.sty}{\usepackage{url}}
{\newcommand{\url}{\text}}

\begin{document}
\vspace*{0.2in}

\begin{flushleft}
{\Large
\textbf\newline{The brain uses renewal points to model random sequences of stimuli. \\
\small \textbf{Dedicated to the memory of Antonio Galves}} 
}
\newline
\\
Fernando A. Najman \textsuperscript{1},
Antonio Galves \textsuperscript{2,\dag},
Marcela Svarc \textsuperscript{3,4},
Claudia D. Vargas\textsuperscript{5,*},
\\
\bigskip
\textbf{1} Instituto de Computação, Universidade Estadual de Campinas, Campinas, Brazil. 
\\
\textbf{2} Instituto de Matem\'atica e Estat\'istica, Universidade de S\~ao Paulo, S\~ao Paulo, Brazil.
\\
\textbf{3} Departamento de Matemática y Ciencias, Universidad de San Andres, Buenos Aires, Argentina.
\\
\textbf{4} CONICET.
\\
\textbf{5}
Instituto de Biof\'isica Carlos Chagas Filho, Universidade Federal do Rio de Janeiro, Rio de Janeiro, Brazil.
\bigskip

%
%



\dag Deceased


*  cdvargas@biof.ufrj.br

\end{flushleft}
\section*{Abstract}
It has been classically conjectured that the brain assigns probabilistic models to sequences of stimuli. An important issue associated with this conjecture is the identification of the classes of models used by the brain to perform this
task. We address this issue by using a new clustering procedure for sets of electroencephalographic (EEG) data recorded from participants exposed to a sequence of auditory stimuli generated by a stochastic chain.
This clustering procedure indicates that the brain uses renewal points in the stochastic sequence of auditory stimuli in order to build a model.

\section*{Author summary}
 A classical conjecture is that the brain is constantly estimating regularities from sequences of events to be able to properly act upon the environment. We propose that, by doing statistics, the brain chooses a model from a class of possible models. Which class of models is used by the brain to encode sequences of events?

  We used an algorithm to generate a sequence of hand claps step by step reproducing a samba-like rhythm. These sequences were generated with stochasticity, were some auditory events were omitted with small probability. We retrieved the regularities of these random sequences of stimuli from EEG data recorded as the participants listened to the samba rhythm.

 To extract the information encoded in the EEG data we introduced a novel procedure for clustering sets of functional data by their relevant statistical features.
 The clusters obtained from the experimental data show that the strong beat of the rhythmic structure is used by the brain to encode the sequence. The strong beat has a remarkable property of separating the sequence into smaller independent blocks. This leads to a natural and economical explanation on how the brain organises the sequence in order to estimate the next event.

\newpage

\section*{Introduction}

It has been proposed that the brain identifies statistical regularities in sequences of stimuli and organises these regularities to be able to classify the stimuli and to make predictions. How the brain proceeds to identify and to classify statistical regularities is still essentially an open question. Some authors use the Bayesian inversion to address this issue \cite{LM03},\cite{MSM15}. Ideas coming from information theory as, for instance, using criteria based on the Kullback-Leibler divergence to select models have been proposed \cite{F05}. Recently Duarte et al. \cite{DFGOV19} and Hernandez et al. \cite{HDOFGV21} proposed the use of probabilistic context tree models \cite{R83,GLO22} to represent the statistical regularities displayed by stochastic sequences of stimuli. In spite of the interest of all these works, each one of them assumes a particular class of models to be used by the brain so as to \textit{organise} \cite{BC81} the set of statistical regularities of sequences of stimuli. However, the identification of the class of models actually used by the brain remains an open question.

We propose to address this question by using a new clustering procedure for sets of functional data. This procedure is applied to sets of electroencephalographic (EEG) data collected while participants are exposed to sequences of auditory stimuli driven by a stochastic chain with memory of variable length.

Our conjecture is that the brain identifies  
\emph{renewal points}\cite{GGGGL12} in the sequence stimuli, and  
uses them to build a model of the sequence. A renewal point is an event in a stochastic stimulus sequence which guarantees that what  
occurs after it is independent of what occurred before it \cite{GGGGL12}.  In this article we describe an  
EEG-based experiment and data analysis methodology that supports this  
conjecture. The analysis relies on the random-projection technique of  
Cuesta-Albertos et al. \cite{CAFR06}.

\section*{Materials and methods}

\subsection*{Experimental protocol}

Nineteen participants were instructed to close their eyes, remain seated, and listen attentively to a random sequence of auditory stimuli presented through earphones. Research was performed in accordance with the relevant guidelines and regulations and approved by the Research Ethics committee of the Institute of Neurology Deolindo Couto at the Federal University of Rio de Janeiro (Plataforma Brasil process number 22047613.2.0000.5261).

The auditory events used as stimuli were strong beats, weak beats, and silent units. The strong and weak beats were recordings of hand claps (spectral frequency range 0.2–15 KHz and maximum duration 200 ms each). The sound files were interrupted at 450 ms with a sharp cutoff. This was done using the software Audacity, version 2.0.5.0 (https://www.audacityteam.org/). The interstimulus interval between two consecutive sound units was always 450 ms.

The sequence of auditory events was chosen in a random fashion for each participant, described in the following section. Electroencephalographic (EEG) signals were recorded during the presentation of the sequences of auditory events. For more details on the experimental protocol see \cite{HDOFGV21}.

\subsection*{Generating the stochastic sequence of auditory stimuli}
Let the elements of the set  $\{0,1,2\}$ represent silences, weak beats, and strong beats, respectively. The sequence $X_n : n = 0,1,\ldots,N$ can be generated symbol by symbol by using the following algorithm. 

These sequences can be generated, in the following manner. Let $\mathcal{V}$ be the set of all participants. For all $v \in \mathcal{V}$ we start with $X^v_0 = 2$. For each $n =  
0,\cdots,N-1$, let $T^v_n$ be the largest $t$ such that $t \leq n$ and  
$X^v_t=2$. Then:
   \begin{itemize}
     \item If $T^v_n$ is $n$ or $n-2$, we set $X^v_{n+1}$ to 1 with  
probability 0.8, or 0 with probability 0.2.
     \item If $T^v_n = n-1$, we set $X^v_{n+1}$ to 1.
     \item If $T^v_n = n-3$, we set $X^v_{n+1}$ to 2.
   \end{itemize}
   This process generates stochastic sequences with a ``samba-like''  
rhythm structure. A sample output is 
$$
\cdots 2 \ 1 \ 0 \ 1 \ 2 \ 0 \ 0 \ 1 \ 2 \ 1 \ 0 \ 0 \ 2 \ 0 \ 0 \ 0 \ 2 \cdots
$$

\noindent Note that the label 2, that represents the strong beat  
stimulus, is a renewal point; that is, for any $n$ and $k$ with $2\leq  
n\leq n+k\leq N$,\[\mathbb{P}(X_n^{n+k} = x_n^{n+k}| X_{n-1}=2,X_0^{n-2}=  
x_0^{n-2}) = \mathbb{P}(X_n^{n+k} = x_n^{n+k}| X_{n-1} = 2);\]where $X_i^j$  
denotes the subsequence $(X_i,X_{i+1},\ldots,X_j)$.






\subsection*{Data acquisition and pre-processing}

In our analysis we used only electrodes $E = \{9, 10, 11, 18, 22, 74, 75, 82\}$, in the standard Geodesic  
numbering of 128 electrode sets \cite{LF05}. The Geodesic amplifier (Geodesic HidroCel GSN 128 EGI, Electrical Geodesic Inc.) was coupled with high input
impedance amplifier (200M$\Omega$, Net Amps, Electrical Geodesics
INC., Eugene, OR, USA). An analogical first order Butterworth band pass filter (0.3-50 Hz) was applied to the signal and the Cz electrode was used as the reference during data acquisition. The signal was acquired with recording frequency of 500 Hz.

In offline processing the data was re-referenced to the average using the EEGLAB package for MATLAB \cite{DM04} and a fourth order Butterworth band pass filter (1-30 Hz) was applied to the signal. For each electrode $e\in  
E$, each participant $v\in \cal V$, and each stimulus sequence index  
$n = 0,1,\ldots,N$, we will denote by $Y^{e,v}_n$ the segment from  
that EEG signal starting $50\,\mathrm{ms}$ before the onset of that  
auditory stimulus and ending at 0.4 seconds after that onset. The  
baseline of each EEG segment $Y^{e,v}_n$ was shifted by subtracting  
from it the average of the signal in the $50\,\mathrm{ms}$ immediately  
preceding the onset of $X_n$.

We further reduced the  
data by combining the electrodes in three subsets $E = \{\mathcal{E}^{\mathrm{RPF}}, \mathcal{E}^{\mathrm{LPF}}, \mathcal{E}^{\mathrm{OCC}}\}$, namely: 
$$ \mathcal{E}^{\mathrm{RPF}} = \{9,10,11\}\quad \mathcal{E}^{\mathrm{LPF}} = \{11, 18, 22\}\quad \mathcal{E}^{\mathrm{OCC}} = 
\{74, 75, 82\} 
$$
For each subset $\mathcal{E} \in E$, each participant  
$v\in \cal V$, and each segment index $n$, we  
denote by $\cal E$ the pointwise average $Y^{\cal E}_n$ of the signals $Y^e_n$ for all  
$e\in\cal E$; namely, $Y^{\cal E}_n = (\sum_{e\in \cal E}  
Y^e_n)/\left|\mathcal{E}\right|$, where $|\mathcal{E}|$ is the cardinality of the set $\mathcal{E}$.

\subsection*{EEG data analysis}

The goal was to retrieve from the EEG data the fingerprints of the putative model that the brain assigns to the sequence of auditory stimuli. This model contains a partition of the set of all possible sequences of past auditory stimuli and an associated family of probability
measures that chooses the next auditory stimulus given the element of the partition to 
which the current past sequence of stimuli belongs to.
\subsubsection*{Clustering sets of EEG segments}
 The first step of  
our analysis was to separately cluster sets of EEG segments of each  
participant $v\in\cal V$. To simplify notation, we will generally omit  
the index $v$ in $Y^{{\cal E},v}_n$ and $X^v_n$ in the remainder of this subsection.

Let $\mathcal{U}$ denote the set of all strings $X_n^{n+2}$ of three labels that may appear  consecutively in any sequence $X_0,X_1,\ldots,X_N$. We denote by $P^*$  
the partition of $\cal U$ determined by the position of the label 2 in  
the substring, or its absence. That is,

$$
P^* = \{\{000,001,101,100\},\{200,210\},\{020,021,120,121\},\{002,012\}\} \ .
$$

\noindent For each participant $v$, each set of electrodes $\mathcal{E}$, each string $u = (u_1,u_2,u_3) \in \mathcal{U}$, and for $n = 2,\ldots,N$ we denote by 
     $$
     \Tilde{\mathcal{Y}}^{\mathcal{E},u} = \{Y^{\mathcal{E}}_n : X^n_{n-2} = u\} \ 
     $$
    the set of all EEG segments recorded during the presentation of stimulus $u_3$ whenever it followed stimuli $u_1$ and $u_2$. We then denote by $\mathcal{Y}^{\mathcal{E},u}$ the set obtained by removing the ten percent most outlying segments from the set $\Tilde{\mathcal{Y}}^{\mathcal{E},u}$, which were ordered using the Fraiman-Muniz functional depth measure \cite{FM01}.

We are assuming that all the EEG segments recorded in the subset $\mathcal{E}$ of electrodes during the presentation of the auditory stimuli indexed by $u_3$ occurring at the end of string $u=(u_1,u_2,u_3)$ have the same stochastic law (or  
\emph{law}, for short \cite{CAFR06}). Let us denote by  $Q^{\mathcal{E},u}$ this law.

Given two strings $u$ and $u'$ and a virtual electrode $\mathcal{E}$, with $u\neq u',$ we define a dissimilarity measure between the laws $Q^{\mathcal{E},u}$ and $Q^{\mathcal{E},u'},$ and estimate it using the samples $\mathcal{Y}^{\mathcal{E},u}$ and $\mathcal{Y}^{\mathcal{E},u'}.$ Given a threshold $\delta \in (0,1]$
    , the dissimilarity measure $\Delta^{\delta}$ between $Q$ and $Q'$ is defined by the formula
    $$
    \Delta^{\delta}(Q,Q')=\int_{L^2[0,T]} \mathbf{1}\{\|F_{\pi_b(Q)}-F_{\pi_b(Q')}\|_{\infty}>\delta\} dP(b),
    $$
where $F_{\pi_b(Q)}$ is the cumulative distribution function of the marginal of $Q$ on $b \in L^2([0,T])$, for $T>0$, and $\mathbf{1}\left\lbrace A \right\rbrace$ is the indicator function of condition $A$, then $\mathbf{1}\left\lbrace A \right\rbrace=1$ if condition $A$ is satisfied  and $\mathbf{1}\left\lbrace A \right\rbrace=0$ otherwise.  

    We estimate the dissimilarity $\Delta^{\delta}(Q^{\mathcal{E},u},Q^{\mathcal{E},u'})$ with the following procedure inspired by the  projective method \cite{CFR07}: 
\begin{enumerate}

    \item Let $B = \left( B(t) 
    : t \in  [-0.05,0.4]\right)$  be a realisation of the Brownian bridge indexed by the time interval $[-0.05,0.4].$ The Brownian bridge $B$ is generated independently from the data set.  For each participant $v$, each subset of electrodes $\mathcal{E} \subset E$ and each string $u$, we denote by $Z^{\mathcal{E},B,u}_n$ the real number defined as the inner product of $Y^{\mathcal{E},u}_n\in \mathcal{Y}^{\mathcal{E},u}$ and $B$. More precisely 
    $$  
    Z^{\mathcal{E},B,u}_n =\int_{-0.05}^{0.4} Y^{\mathcal{E},u}_n(t_n+s)B(s)ds,
    $$
    where $t_n= n\times0.4-0.05$ is the starting time of the EEG segment $Y^{\mathcal{E},u}_n$. We denote by $Z^{\mathcal{E},B,u}_n $ the \textit{``projection''} of the segment $Y^{\mathcal{E},u}_n$ in the \textit{``direction''} $B.$
    \item Denote $\mathcal{Z}^{\mathcal{E},B,u}$ the set of projections of the EEG segments belonging to $\mathcal{Y}^{\mathcal{E},u}$ in the direction $B.$ To simplify the notation in the following we omit $\mathcal{E}$ unless otherwise noted.
    \item Let $B_1,\dots, B_m$ be a sequence of independent realizations of the Brownian Bridge indexed by the time interval $[-0.05,0.4]$ and generated independently of the data set. For each pair of strings $u \in \mathcal{U}$ and $u'\in\mathcal{U}$, $u \neq u'$, we define the $\widehat{\Delta}_{m,n}^{\delta}(Q^u,Q^{u'})$ empirical dissimilarity as

$$
\widehat{\Delta}_{m,n}^{\delta}(Q^u,Q^{u'})= \frac{1}{m}\sum_{j=1}^m
\mathbf{1} \{ \|\widehat{F}^{u,B_j}_{n}-\widehat{F}^{u',B_j}_{n} \|_{\infty}  >\delta\},
$$

    where
    $$
    \widehat{F}^{u,B_j}(t)= \frac{1}{|\mathcal{Z}^{u,B_j}|} \sum_{z \in \mathcal{Z}^{u,B_j}} \mathbf{1}\{z\leq t \} 
    $$
    and
     $$
    \widehat{F}^{u',B_j}(t)= \frac{1}{|\mathcal{Z}^{u,B_j}|} \sum_{z \in \mathcal{Z}^{u',B_j}} \mathbf{1}\{z\leq t \}. 
    $$
    \end{enumerate}

\begin{figure}
    \centering
    \includegraphics[scale=0.25]{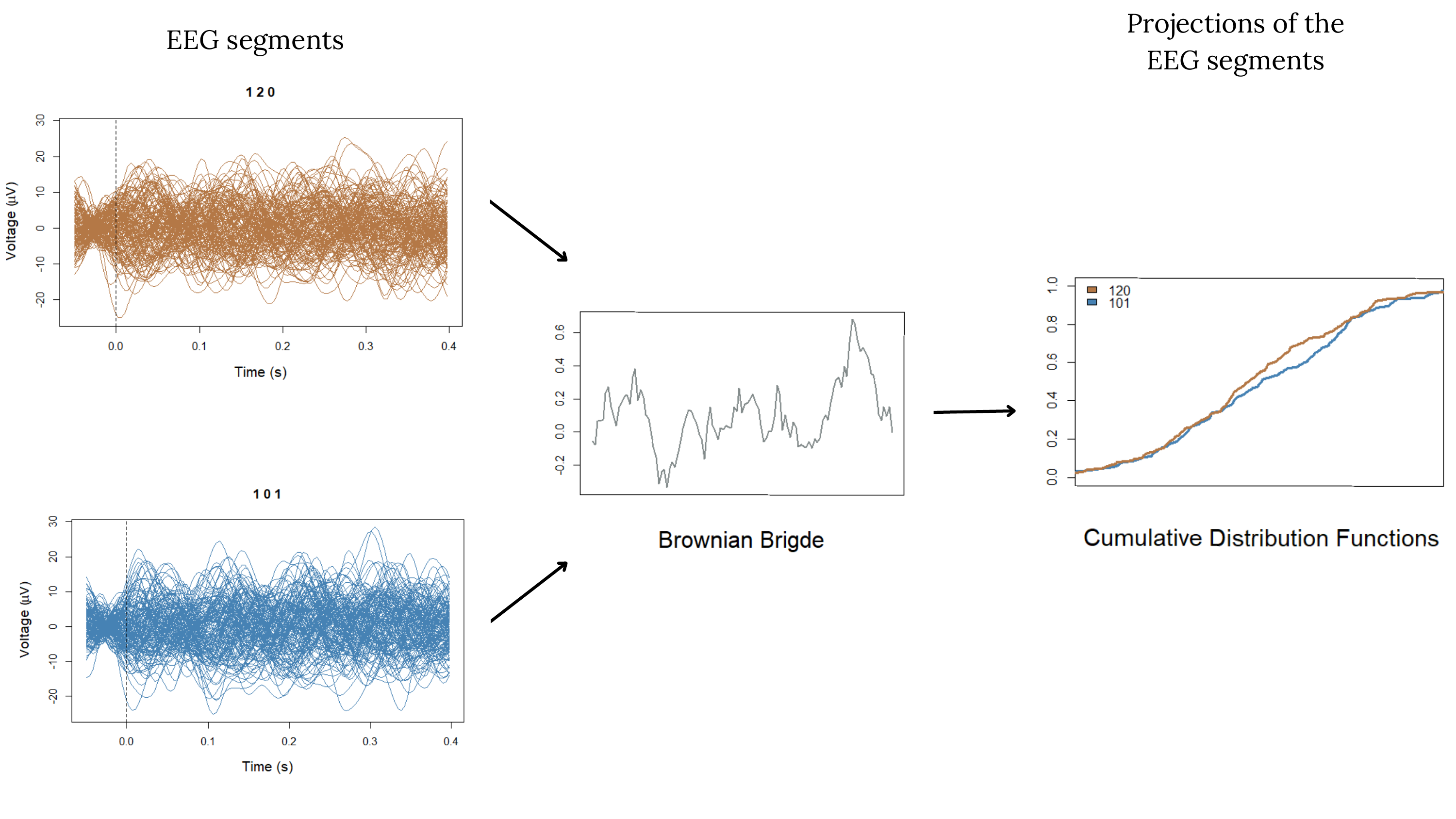}
    \caption{Example of the projective method applied to sets of EEG segments (120 and 101) for a given participant and electrode. All segments of both sets are projected into the same independent realization of a Brownian Bridge. For each set, we obtain a new set of real numbers. We can then compare the distribution of the two sets of real numbers obtained. Here we represent these distributions by their cumulative distribution functions and histograms.}
    \label{fig:projections}
\end{figure}

This procedure is depicted in Fig. \ref{fig:projections}.

        For each participant $v$ and each subset $\mathcal{E}\subset E$, using the empirical dissimilarity matrix $\widehat{\Delta}_{m,n}^{\delta}$, we build a dendrogram with the sets $\{\mathcal{Y}^{u} : u \in \mathcal{U}\}$ of EEG segments by using the hierarchical clustering procedure with complete linkage with $m = 5000$ and $\delta = 276/5000$, as in \cite{HDOFGV21}. Using a fixed threshold we finally obtain a partition of the sets $\{\mathcal{Y}^{u} : u \in   \mathcal{U}\}$ of EEG segments, which we denote by $\mathcal{C}$.

\subsubsection*{Summarizing partitions obtained from each participant}
For each fixed set of electrodes $\mathcal{E} \subset E$ the procedure we have just described produces a partition of the sets $\{\mathcal{Y}^{u} : u 
    \in 
    \mathcal{U}\}$ for each participant $v$. For each $v \in \mathcal{V}$, we denote as $\mathcal{C}^v$ the partition obtained with the procedure described above. Given two strings $u$ and $u'$, we say that $u \overset{v}{\sim} u' $, if the sets $\mathcal{Y}^u$ and $\mathcal{Y}^{u'}$ belong to the same cluster of the partition $\mathcal{C}^v$.

    To summarise the results across participants we then performed an aggregation consensus clustering. By an aggregation consensus clustering we mean the following. We define a new dissimilarity matrix $\eta$ for the elements of $\mathcal{U}$ with entries defined as follows. For each pair $(u,u') \in \mathcal{U}^2$,
   
    $$
    \eta(u,u') = \sum_{(v,v') \in \mathcal{V}^2}\mathbf{1}\{u\overset{v}{\sim} u' \}\mathbf{1}\{u\overset{v'}{\sim}u' \} \ .
    $$

\begin{figure}
    \centering
    \includegraphics[scale=0.2]{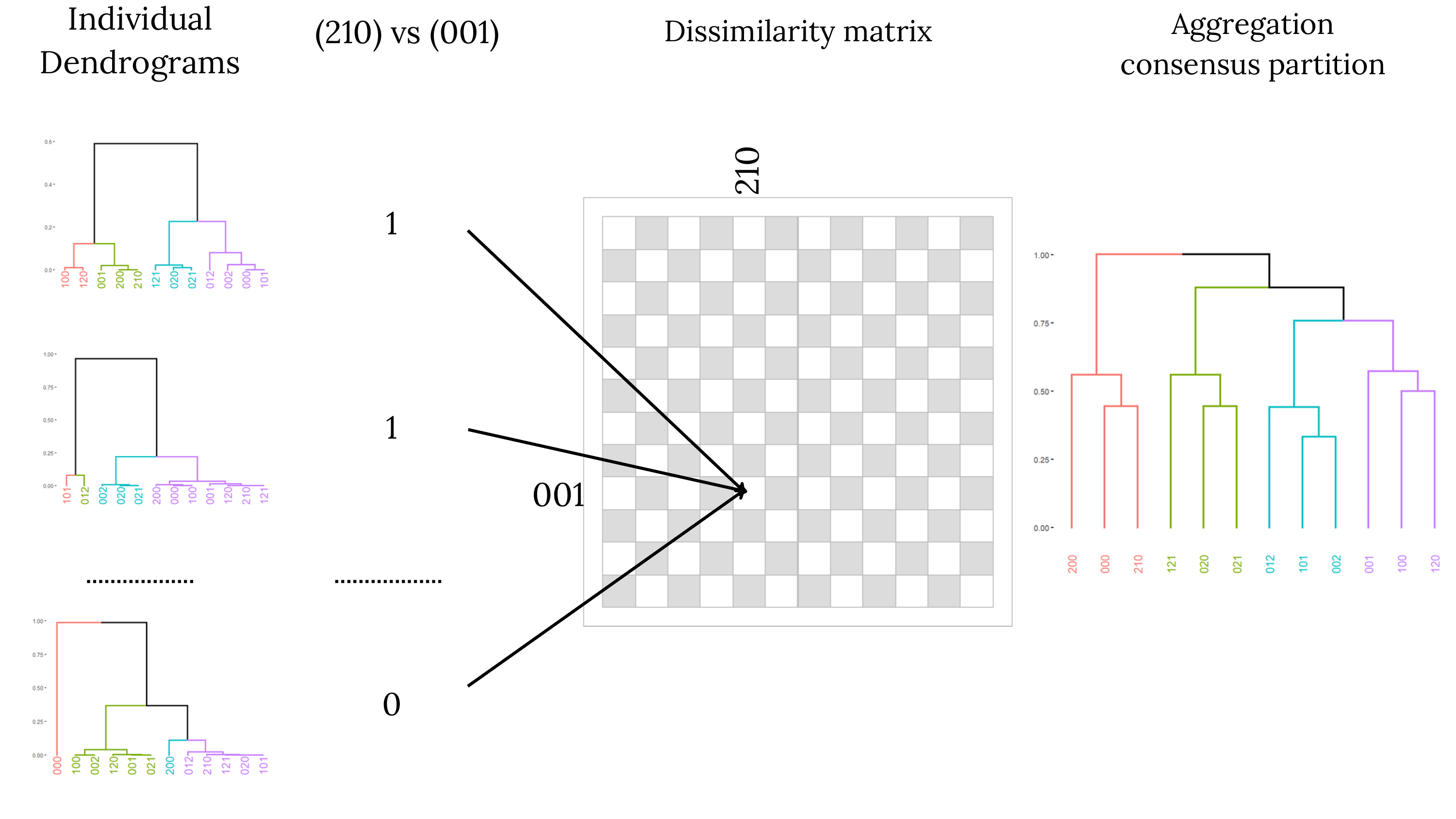}
    \caption{Procedure for the construction of a consensus dendrogram. For each individual dendrogram obtained with the method described in the section 'Clustering sets of EEG data' and any pair of strings in $\mathcal{U}$, we assign to each dendrogram 1 if the two strings are in the same cluster, and 0 otherwise. We take the average of these labels as the dissimilarity between the two strings. Finally, using this dissimilarity matrix, we construct the consensus dendrogram.}
    \label{fig:consensusclus}
\end{figure}

    Once we have the dissimilarity matrix $\eta$, we obtain a new dendrogram by using a hierarchical clustering procedure on the set $\mathcal{U}$  with the Ward linkage \cite{MAET15}, \cite{W63}. Finally we select the partition with four clusters. We call the outputs of this procedure as  \textit{consensus dendrogram} and \textit{consensus partition}. We denote the consensus partition by $P^{\mathcal{E}}$. This process is shown in Fig. \ref{fig:consensusclus}. All codes are accessible at \href{https://github.com/fanajman/The-brain-uses-renewal-points-to-model-random-sequences-of-stimuli.git}{https://github.com/fanajman/The-brain-uses-renewal-points-to-model-random-sequences-of-stimuli.git} and the data is acessible at \href{https://neuromat.numec.prp.usp.br/neuromatdb/EEGretrieving/}{ https://neuromat.numec.prp.usp.br/neuromatdb/EEGretrieving/}. 

\subsubsection*{Statistical significance of the consensus partitions}\label{subsec:statsignificance}

To test the similarity between the  consensus partitions and $P^*$, we conducted a numerical estimation of the probability of retrieving these partitions under the following null hypothesis.

Let $R_i:1,\cdots,10^6$ be random symmetric matrices in $\mathbb{R}^{12\times12}$ where the entries of the upper diagonal of each matrix are independent uniform variables assuming values in the $[0,1]$ interval. For each random matrix $R_i$,  we retrieve a partition $P_i$ with four elements using the hierarchical clustering procedure with the Ward linkage. 

Given two partitions $P_1$ and $P_2$, we denote as  $ARI(P_1,P_2)$ the value obtained by employing the adjusted Rand index \cite{HA85}, which is a measure of similarity between the between the pair $(P_1,P_2)$. This statistic ranges in the $[-1,1]$ interval, returning $1$ for two identical partitions.

For each set of electrodes $\mathcal{E} \subset E$ we take the proportion 
    $$
    \hat{p}(\mathcal{E}) = \frac{1}{10^6} \sum_{i=1}^{10^6} \mathbf{1}\{ARI(P_i,P^*)\geq ARI(P^*,P^{\mathcal{E}})\}
    $$
    as the estimated probability of finding a partition at least as similar to the partition $P^*$ defined by the renewal point, as described in subsection 'Clustering of EEG segments per participant'.

\section*{Results}

\begin{figure}[h!]
    \centering
    \includegraphics[scale=0.4]{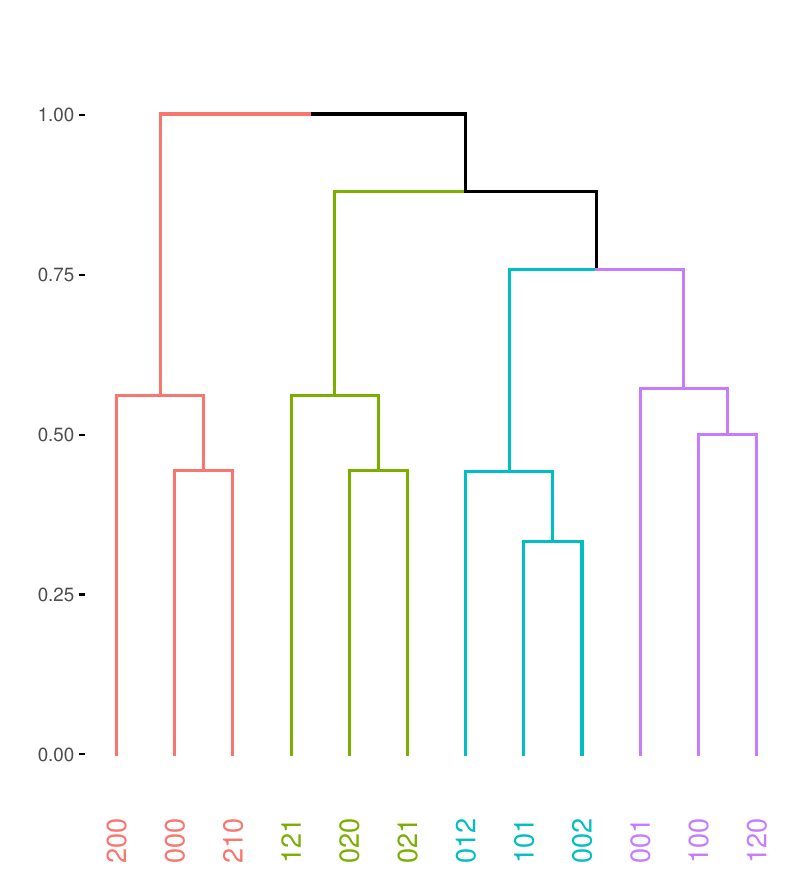}
    \caption{Partition and associated dendrogram retrieved from the EEG segments recorded in $\mathcal{E}^{RPF}$. The black line represents a partition of the strings with more than one of the final clusters. The other colours represent the clusters of $P^{\mathcal{E}_{RPF}}$.}
    \label{fig:dendroRPF}
\end{figure}

\begin{figure}[h!]
    \centering
    \includegraphics[scale=0.4]{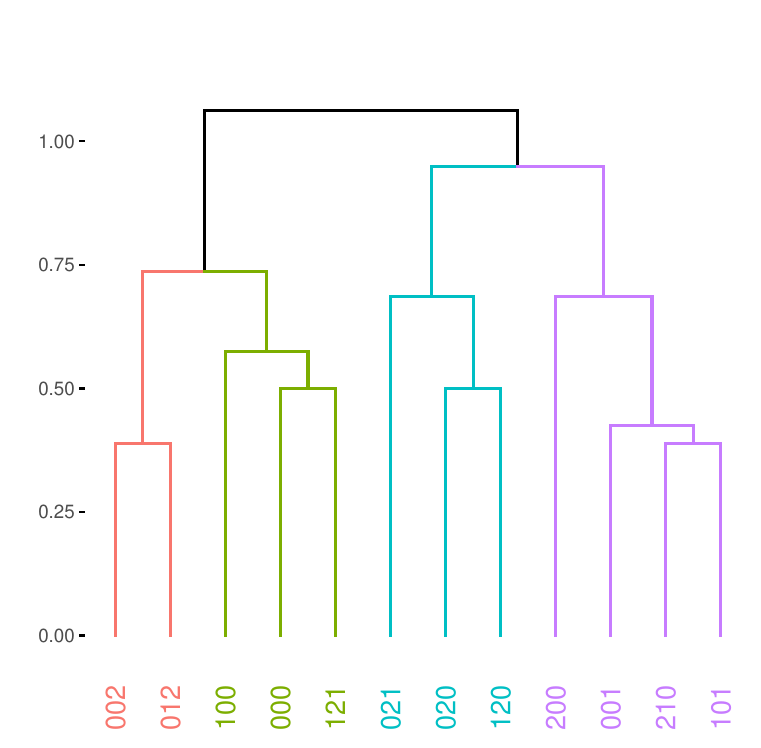}
    \caption{Partition and associated dendrogram retrieved from the EEG segments recorded in $\mathcal{E}^{LFP}$. The black line represents a partition of the strings with more than one of the final clusters. The other colours represent the clusters of $P^{\mathcal{E}_{LPF}}$.}
    \label{fig:dendroLPF}
\end{figure}

\begin{figure}[h]
    \centering
    \includegraphics[scale=0.4]{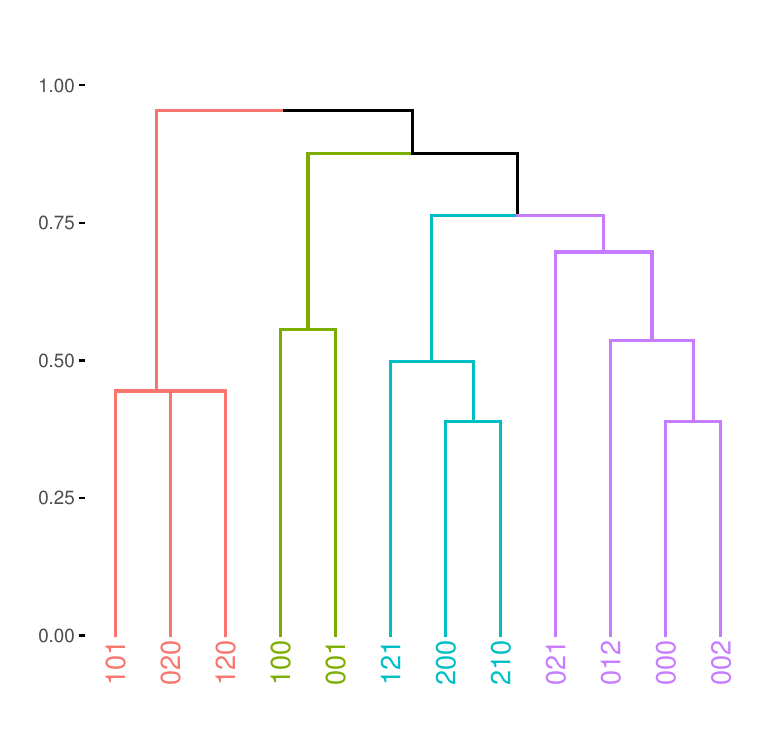}
    \caption{Partition and associated dendrogram retrieved from the EEG segments recorded in $\mathcal{E}^{OCC}$. The black line represents a partition of the strings with more than one of the final clusters. The other colours represent the clusters of $P^{\mathcal{E}_{OCC}}$.}
    \label{fig:dendroOCC}
\end{figure}

The consensus partitions and associated consensus dendrograms obtained from the sets $(\mathcal{E}^{RPF},\mathcal{E}^{LPF},\mathcal{E}^{OCC})$ of electrodes are shown in Figs. \ref{fig:dendroRPF}, \ref{fig:dendroLPF} and \ref{fig:dendroOCC}.

Fig. \ref{fig:dendroRPF} shows the consensus partition and associated consensus dendrogram retrieved from the subset $\mathcal{E}^{RPF}$. In this partition, 9 out of the 12 strings were classified as expected. The partition obtained contains the following clusters.
\begin{itemize}
\item A cluster with the strings $012$ and $002$, the two strings that end in the strong beat with probability $1$. This cluster also contains the string $101$.
 \item A cluster with the strings $021$, $020$ and $121$. These three strings end one position after the strong beat.
    \item A cluster with the strings $200$ and $210$. These are the two strings ending in a silent unit which occurs with probability $1$. This cluster also contains the string $000$.
   \item A cluster with the strings $001$ and $100$. These strings end four positions after the strong beat. This cluster also contains the string $120$.
\end{itemize}

Figure \ref{fig:dendroLPF} shows the consensus partition and associated consensus dendrogram retrieved from the set $\mathcal{E}^{LPF}$. This partition is similar to the right prefrontal electrodes in the sense that
\begin{itemize}
    \item The two strings ending in the strong beat were assigned to the same cluster.
    \item The two strings ending in a silent unit with probability one were assigned to the same cluster.
    \item The consensus partition contains a cluster consisting exclusively of three out of four strings which end one position after the strong beat.
\end{itemize}

\begin{figure}[h]
    \centering
    \includegraphics[scale=0.2]{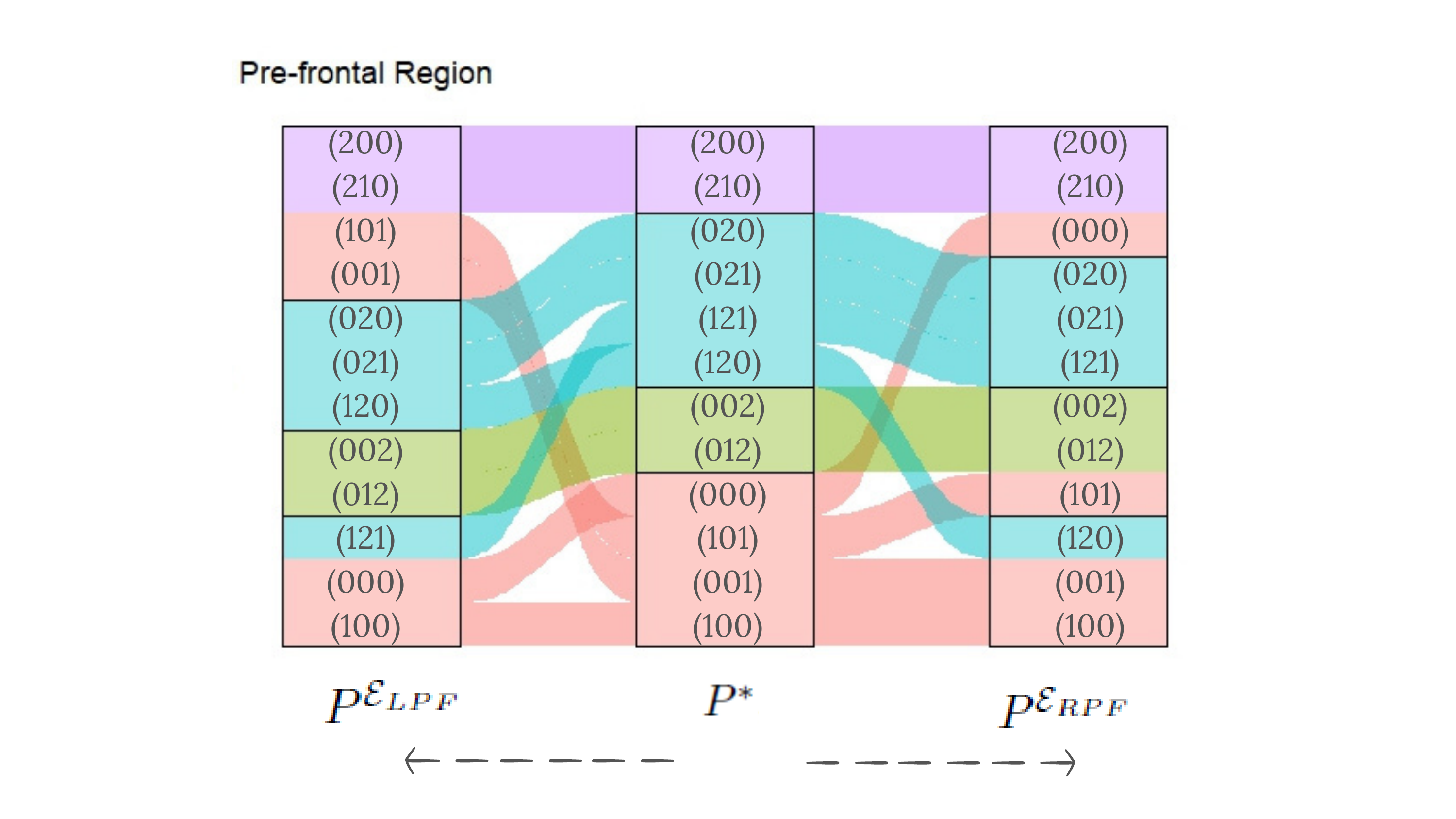}
    \caption{Sankey diagram showing the matching between $P^{\mathcal{E}_{LPF}}$, $P^*$,  and $P^{\mathcal{E}_{LPF}}$. The middle column presents the $P^*$ partition. The colours represent the position of the strong beat in the string.}
    \label{fig:partitioncompar}
\end{figure}

 The Sankey diagram in Fig. \ref{fig:partitioncompar} exhibits the matching between $P^*$, $P^{\mathcal{E}_{RPF}}$ and $P^{\mathcal{E}_{LPF}}$.

Fig. \ref{fig:dendroOCC} shows the partition and associated dendrogram obtained from the occipital electrodes, chosen as a control region. As in the prefrontal partitions, the two strings ending in the strong beat were assigned to the same cluster and the two strings ending in the silent unit that occurs with probability 1 were also assigned to the same cluster. The strings ending in the constituent silent unit were assigned to a cluster which also contains two strings ending one step before the strong beat. No cluster containing exclusively strings ending one step after the strong beat were obtained.

For each partition retrieved from a subset of electrodes, the probability of finding a partition at least as similar as the partition expected was numerically estimated with the method described in the 'Statistical significance of the consensus partitions' section. In blue are the p-values smaller than the usual 0.05 criterion used to consider the effect significant. 
\begin{itemize}
    \item Right prefrontal region: \textcolor{blue}{0.0174}.
    \item Left prefrontal region: \textcolor{blue}{0.0096}.
    \item Occipital region: 0.1998.
\end{itemize}

\section*{Discussion}
 The consensus partitions retrieved from prefrontal EEG data show significant similarity to the partition expected by the occurrence of the renewal points. These results agree with our conjecture that the brain uses the strong beat represented by the symbol $2$ to model the stimuli sequence. More precisely, the brain selects a partition of the past of the stimuli sequence with only four clusters, this partition being defined by last occurrence of the strong beat.

The use of the recurrent occurrence of the strong beat to model the stimuli sequence shows that the brain is able to find a model of the sequence, that is, to identify the structural aspects of the stimuli sequence to accurately estimate the next event amidst the stochasticity.

  A partition of the past of the sequence of stimuli with only four elements is an economical representation of the sequence, since this is the smallest partition of the past which allows generating the stimuli sequence step-by-step. This compression scheme is more efficient than the context tree model proposed in \cite{HDOFGV21}, since this partition has four elements, in contrast with the eight elements of the context tree.

 The occurrence of the strong beat regenerates the sequence, meaning that the stimuli sequence can be partitioned in independent blocks demarcated by the occurrence of each strong beat. It is a well known fact that participants can perceive sequences of stimuli as independent blocks, a phenomenon called chunking (for a  review see \cite{DMWWP15}). Therefore our conjecture gives us a natural explanation for this phenomenon. How to estimate if an event is a renewal point is in itself an open statistical question which has only recently started being investigated \cite{FG23}. 

 In \cite{LK94} it is conjectured that the activity of some neuron populations synchronizes with temporally regular occurrences in the auditory stimuli, a phenomenon sometimes called entrainment. This is a possible mechanism used by the brain to model the stimuli sequence, considering that in our experiment the isochronous occurrence of a strong beat every four auditory units is a recurrent regularity. The entrainment phenomenon has been shown to explain chunking of continuous auditory stimuli in the temporal scale of 150–300 ms in the temporal cortex \cite{TTDP18}. Our results indicate that the prefrontal cortex uses the occurrence of the strong beat to chunk the auditory sequence in larger timescales, with seconds between the occurrence of each strong beat.

\section*{Conclusion}
We show evidence that the brain uses the renewal points of the stochastic sequence of auditory stimuli in order to select a model of a random sequence of stimuli.

\section*{Acknowledgments}
This work is part of the activities of FAPESP Research, Innovation and Dissemination Center for Neuromathematics (grant $\#$ 2013/ 07699-0 , S.Paulo Research Foundation (FAPESP). This work is supported by CAPES (88882.377124/2019-01) and FAPESP (2022/00784-0) grants. A.G and C.D.V. were partially supported by CNPq fellowships (grants 314836/2021-7 and 310397/2021-9) This article is also supported by FAPERJ ( $\#$ CNE 202.785/2018 and $\#$ E- 26/010.002418/2019), and FINEP ( $\#$ 18.569-8) grants. 

The authors acknowledge the hospitality of the Institut Henri Poincar\'e (LabEx CARMIN ANR-10-LABX-59-01) where part of this work was written and Jorge Stolfi for his insightful comments on the manuscript.

%
%
%



\begin{thebibliography}{99}

\bibitem{LM03}
Lee, T. S., \& Mumford, D. (2003). Hierarchical Bayesian inference in the visual cortex. JOSA A, 20(7), 1434-1448.

\bibitem{MSM15}
Meyniel, F., Sigman, M., \& Mainen, Z. F. (2015). Confidence as Bayesian probability: From neural origins to behavior. Neuron, 88(1), 78-92.

\bibitem{F05}
Friston, K. (2005). A theory of cortical responses. Philosophical transactions of the Royal Society B: Biological sciences, 360(1456), 815-836.



\bibitem{DFGOV19} Duarte, A., Fraiman, R., Galves, A., Ost, G. \& Vargas, C. D. (2019). Retrieving a Context Tree from EEG Data. Mathematics, 7(5), 427. 

\bibitem{HDOFGV21} 
Hernández, N., Duarte, A., Ost, G., Fraiman, R., Galves, A., \& Vargas, C. D. (2021). Retrieving the structure of probabilistic sequences of auditory stimuli from EEG data. Scientific Reports, 11(1), 3520.


\bibitem{R83}
Rissanen, J. (1983). A universal data compression system. IEEE Transactions on information theory, 29(5), 656-664.


\bibitem{GLO22}
Galves, A., Leonardi, F. G., \& Ost, G. (2022). Statistical model selection for stochastic systems with applications to bioinformatics, linguistics and neurobiology.
\url{ https://impa.br/wp-content/uploads/2022/01/33CBM15-eBook.pdf}

\bibitem{BC81}
Bregman, A. S., \& Campbell, J. (1971). Primary auditory stream segregation and perception of order in rapid sequences of tones. Journal of experimental psychology, 89(2), 244.


\bibitem{GGGGL12}
Galves, A., Galves, C., García, J. E., Garcia, N. L., \& Leonardi, F. (2012). Context tree selection and linguistic rhythm retrieval from written texts. Annals of Applied Statistics 6(1): 186-209.

\bibitem{LF05}
Luu, P., \& Ferree, T. (2005). Determination of the HydroCel Geodesic Sensor Nets’ average electrode positions and their 10–10 international equivalents. Inc, Technical Note, 1(11), 7.

\bibitem{DM04}
Delorme, A., \& Makeig, S. (2004). EEGLAB: an open source toolbox for analysis of single-trial EEG dynamics including independent component analysis. Journal of neuroscience methods, 134(1), 9-21.

\bibitem{FM01}
Fraiman, R., \& Muniz, G. (2001). Trimmed means for functional data. Test, 10, 419-440.

\bibitem{CAFR06}
Cuesta-Albertos, J. A., Fraiman, R., \& Ransford, T. (2006). Random projections and goodness-of-fit tests in infinite-dimensional spaces. Bulletin of the Brazilian Mathematical Society, 37(4), 477-501.

\bibitem{CFR07} Cuesta-Albertos, J.A., Fraiman, R. \& Ransford, T. (2007) A Sharp Form of the Cramér–Wold Theorem. J Theor Probab 20, 201–209.

\bibitem{MAET15}
Miyamoto, S., Abe, R., Endo, Y., \& Takeshita, J. I. (2015, November). Ward method of hierarchical clustering for non-Euclidean similarity measures. In 2015 7th International Conference of Soft Computing and Pattern Recognition (SoCPaR) (pp. 60-63). IEEE.

\bibitem{W63}
Ward Jr, J. H. (1963). Hierarchical grouping to optimize an objective function. Journal of the American statistical association, 58(301), 236-244.

\bibitem{HA85}
Hubert, L., \& Arabie, P. (1985). Comparing partitions. Journal of classification, 2, 193-218.

\bibitem{DMWWP15}
Dehaene, S., Meyniel, F., Wacongne, C., Wang, L., \& Pallier, C. (2015). The neural representation of sequences: from transition probabilities to algebraic patterns and linguistic trees. Neuron, 88(1), 2-19.

\bibitem{FG23}
Freguglia, V., \& Garcia, N. L. (2023). Detecting renewal states in chains of variable length via intrinsic Bayes factors. Statistics and Computing, 33(1), 21.

\bibitem{LK94}
Large, E. W., \& Kolen, J. F. (1994). Resonance and the perception of musical meter. Connection science, 6(2-3), 177-208.



\bibitem{TTDP18}
Teng, X., Tian, X., Doelling, K., \& Poeppel, D. (2018). Theta band oscillations reflect more than entrainment: behavioral and neural evidence demonstrates an active chunking process. European Journal of Neuroscience, 48(8), 2770-2782.






















\end{thebibliography}
\end{document}